\documentclass[prb,superscriptaddress]{revtex4-1}
\usepackage{graphicx}
\usepackage{picture}
\usepackage{xcolor}
\usepackage{color}
\usepackage{tikz}

\begin{document}

\author{H. F. Fotso}
\affiliation{Department of Physics, University at Albany (SUNY),  Albany, New York 12222, USA}
\author{J. K. Freericks}
\affiliation{Department of Physics, Georgetown University, 37th and O Sts. NW, Washington, DC 20057, USA}

%\title[Nonequilibrium Dynamics of a Field-Driven System]{Characterizing the Nonequilibrium Dynamics of Field-Driven Correlated Quantum Systems} 

\title{Characterizing the Nonequilibrium Dynamics of Field-Driven Correlated Quantum Systems} 

\begin{abstract}
Recent experimental advances in ultrafast phenomena have triggered renewed interest in the dynamics of correlated quantum systems away from equilibrium. 
We review nonequilibrium dynamical mean-field theory studies of both the transient and the steady states of a DC field-driven correlated quantum system. In particular, we focus on the nonequilibrium behavior and how it relates to the fluctuation-dissipation theorem. The fluctuation-dissipation theorem emerges as an indicator for how the system thermalizes and for how it reaches a steady state. When the system thermalizes in an infinite temperature steady state it can pass through a succession of quasi-thermal states that approximately obey the fluctuation-dissipation theorem. We also discuss the Wigner distribution and what its evolution tells us about the nonequilibrium many-body problem.\\

\end{abstract}

{
\let\clearpage\relax
\maketitle
}

\section{Introduction}
\label{sec: Introduction}

Strongly correlated systems include some of the most technologically promising materials of our time. The same quantum-mechanical complexity behind their most intriguing properties also renders them challenging to study. Much of the interest is motivated by recent experimental successes. For example, trapping and manipulating ultracold atomic gases in optical lattices provides a new platform for the controlled examination of strongly correlated systems in and out equilibrium~\cite{ MGreinerBlochNature,BlochDalibardZwerger}. In electronics, device miniaturization leads to the creation of nanoscale devices in which electrons experience strong electric fields and thus cannot be well approximated by linear-response theories~\cite{MeirWingreen_PRL1992, CaroliNozieres_JPhysC1971}. Additionally, pump-probe spectroscopy offers a new avenue for the exploration of available electronic states in correlated materials~\cite{Perfetti_et_al}. These advances have revived interest in the fundamental behavior of quantum systems away from equilibrium.

There are many unanswered questions. How does an equilibrium quantum system that is suddenly driven out of equilibrium, subsequently relax to a thermal state~\cite{Deutsch1991, RigolNature2008, Srednicki1994, thermalization}?
What governs the nonequilibrium driving of a system into a metastable state that is not found among known equilibrium phases~\cite{NonEqPhaseTransition1, NonEqPhaseTransition2}? Answering these questions requires an accurate theoretical approach. Nonequilibrium dynamical mean-field theory~\cite{DMFT_noneq} is a powerful tool to address these pressing questions. 

In equilibrium, dynamical mean-field theory (DMFT)~\cite{DMFT, DMFT_2, DMFT_3, DMFT_FK} treats spatial correlations in a mean-field fashion, while treating temporal correlations exactly. It is one of the most commonly used and successful methods for studying strongly correlated systems. It has also been extended to  nonequilibrium~\cite{DMFT_noneq, FK_NonEq_DMFT08, DMFT_noneq_Aoki}.

Field-driven correlated systems display  nontrivial dynamics. When a DC electric field is applied, the system responds by creating an electric current that flows. That current also leads to Joule heating, which, if left unchecked, will heat the system to infinite temperature (where the current vanishes and the heating ultimately stops). The heating can also stop prematurely, with the system reaching a metastable (or possibly even thermal) state.  Indeed, several scenarios for this relaxation process can occur depending on the specific details of the system~\cite{thermalization}. In many of these situations, the fluctuation-dissipation theorem is no longer rigorously valid. However, it still remains an important concept in nonequilibrium, because it can be used to tell us when a system settles into a thermal steady state~\cite{steadyStateJoura_1, steadyStateJoura_2, steadyState_Aoki, steadyState_Camille_FDT}. We review nonequilibrium DMFT manifestations of these properties in a mixture of heavy and light fermionic particles described by the Falicov-Kimball model~\cite{FalicovKimball, AtesZiegler} that starts in equilibrium and then has a DC electric field suddenly turned on.

The review is organized as follows. In section \ref{sec:NonEqGreensFunctions}, we present a brief description of the nonequilibrium Green's functions and introduce the Falicov-Kimball model as it relates to the Hubbard model. In section \ref{sec:NonEqDMFT}, we describe the nonequilibrium DMFT solution first for the transient and then for the steady state. In section \ref{sec:NonEqDynamicsResults}, we present a set of results that include manifestations of the fluctuation-dissipation theorem and provide insights into the rich dynamics of the field-driven strongly correlated quantum system.

\section{Nonequilibrium Green's Functions}
\label{sec:NonEqGreensFunctions}
\begin{figure}[htbp]
\begin{center}
{\includegraphics[width=5.50cm, height=3.50cm]{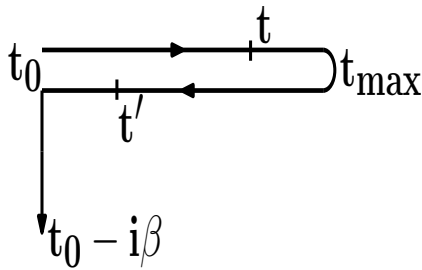}} 
\caption{Kadanoff-Baym-Keldysh contour. The arrows indicate the direction of time ordering. In this example, $t$ occurs before $t'$ on the contour, even though $t>t'$, when thought of as real numbers. 
\label{fig:KeldyshContour}
}
\end{center}
\end{figure}

The many-body formalism for the nonequilibrium problem is similar to that of the equilibrium problem. All essential requirements for perturbation theory still hold away from equilibrium, except for the identification of the state at long times being identical to the state for the earliest times.  Instead, it becomes necessary to evolve the system according to the Heisenberg representation for operators: start from the distant past, evolve forward to the time of physical interest and then evolve backward to the distant past (because of the Hermitian conjugate of the evolution operator in the Heisenberg representation)~\cite{Keldysh64_65,BaymKadanoff62}. This gives rise to the Kadanoff-Baym-Keldysh contour illustrated in figure~\ref{fig:KeldyshContour}; the contour-ordered Green's functions has time ordering performed with respect to the advance along the contour. It is defined via
\begin{equation}
    G^c_{k, \sigma}(t, t')  =  \theta_c(t, t')G^>_{k, \sigma}(t, t') + \theta_c(t', t)G^<_{k, \sigma}(t, t'), \;\;\;\;\label{eq:GContour} 
\end{equation}
where $\theta_c(t,t')$ is the contour-ordered Heaviside function, which orders time with respect to the contour. It is equal to $1$ if $t$ is ahead of $t'$ on the contour and is equal to $0$ if $t$ is behind $t'$ on the contour.
Hence, the nonequilibrium formalism automatically includes several different Green's functions depending on the location of each time argument on the contour. The lesser ($t$ on lower, $t'$ on upper), greater ($t$ on upper, $t'$ on lower), time-ordered ($t$ and $t'$ on upper), anti-time-ordered ($t$ and $t'$ on lower) Green's functions are respectively defined by the following when both time arguments of the contour-ordered Green's function are {\it real}:
\begin{eqnarray}
 G^<_{k, \sigma}(t, t') & = &  i\langle c^{\dagger}_{k \sigma}(t') c_{k \sigma}^{\phantom\dagger}(t)\rangle \label{eq:GLesser}  \\
 G^>_{k, \sigma}(t, t') & = & -i\langle c_{k \sigma}^{\phantom\dagger}(t) c^{\dagger}_{k\sigma}(t')\rangle \label{eq:GGreater} \\
 G^t_{k, \sigma}(t, t') & = & \theta(t-t')G^>_{k, \sigma}(t, t') + \theta(t'-t)G^<_{k, \sigma}(t, t') \;\;\;\;\label{eq:GTOrdered} \\
 G^{\bar{t}}_{k, \sigma}(t, t') & = &  \theta(t'-t)G^>_{k, \sigma}(t, t') + \theta(t-t')G^<_{k, \sigma}(t, t'). \;\;\;\;\label{eq:GAntiTOrdered} 
 \end{eqnarray}
 From these Green's functions, we can construct the so-called retarded and advanced Green's functions via
 \begin{eqnarray}
 G^R_{k, \sigma}(t,t') & = & -i\theta(t-t')\langle \{c_{k\sigma}^{\phantom\dagger}(t), c^{\dagger}_{k\sigma}(t')\}\rangle \label{eq:GRetarded}\\
 G^A_{k, \sigma}(t, t') & = & i\theta(t'-t)\langle \{c_{k\sigma}^{\phantom\dagger}(t), c^{\dagger}_{k\sigma}(t')\}\rangle.\label{eq:GAdvanced}
\end{eqnarray}
Here $c^{\dagger}_{k \sigma}(t)$ and $c_{k \sigma}(t)$ are respectively the Heisenberg representation of the creation and the destruction operators for an electron of momentum $k$ and spin $\sigma$ at time $t$; $\theta$ is the ordinary Heaviside function, {\it i.~e.},~$\theta(t-t') = 0$ if $t<t'$ and $\theta(t, t') = 1$ otherwise; $\{\mathcal{A}, \mathcal{B}\}$ is the anticommutator of operators $\mathcal{A}$ and $\mathcal{B}$. The symbol $\langle \mathcal{O} \rangle$ is the expectation value taken of the operator $\mathcal{O}$ with respect to the initial thermal state:
\begin{equation}
    \langle \mathcal{O} \rangle=\frac{{\rm Tr} e^{-\beta\mathcal{H}(t_{\rm min})}\mathcal{O}}{{\rm Tr} e^{-\beta\mathcal{H}(t_{\rm min})}},
\end{equation}
where $\mathcal{H}(t_{\rm min})$ is the initial Hamiltonian before any field is turned on.
Several additional Green's functions are included in the formalism when at least one time argument is on the imaginary spur of the contour. These include the imaginary-time (or Matsubara) Green's functions with both times on the imaginary time vertical branch and the mixed-time Green's functions where one of the times is on the horizontal branch while the other is on the vertical branch (formulas not explicitly shown here). The nonequilibrium many-body formalism works directly with the contour-ordered Green's functions~\cite{DMFT_noneq}. All other Green's functions can then be extracted from it. 

Note that these Green's functions are not all independent and are related by various identities and symmetries. One choice for two independent ones is  the lesser and the retarded Green's functions, $G^<$ and $G^R$. These determine most physical quantities. In particular, the retarded Green's function is related to the quantum states while the lesser Green's function is directly related to how those states are occupied by fermions.

We are interested in the dynamics of a strongly correlated system that starts initially in thermal equilibrium and then is driven out of equilibrium by turning on an external DC electric field. We describe how the fluctuation-dissipation theorem manifests itself and how it can be utilized in characterizing relaxation through transient states to the (long-time) steady-state.

As an example, we consider a generalized Hubbard model for strongly correlated systems with spin-dependent hopping: 
\begin{equation}
  \mathcal{H}_{eq}   =   - \sum\limits_{\langle ij \rangle, \sigma} J_{ij\sigma} \left( c^{\dagger}_{i\sigma}c_{j\sigma}^{\phantom\dagger} + h.c.\right)
           -  \mu \sum\limits_{i\sigma}c^{\dagger}_{i\sigma}c_{i\sigma}^{\phantom\dagger} +  U\sum\limits_{i}n_{i\sigma}n_{i\bar{\sigma}}.
  \label{eq:HamiltonianGeneric}        
\end{equation}
Here $\langle ij \rangle$ represents unique nearest-neighbor pairs for sites $i$ and $j$; $J_{ij\sigma}$ is the nearest-neighbor hopping integral; $\mu$ is the chemical potential; $n_{i \sigma} = c^{\dagger}_{i\sigma}c_{i\sigma}^{\phantom\dagger}$ is the number operator for electrons of spin $\sigma$ at site $i$; and $U$ is the on-site repulsion for doubly occupied sites.

When $J_{ij\sigma} = J_{ij\bar{\sigma}} = J$, we obtain the conventional Hubbard model, which is believed to contain the essential elements for high-temperature superconductivity in the cuprates and for this reason has been the subject of intense research activity. Despite its deceptive simplicity, the model remains unsolved except in one dimension~\cite{LiebWu_1968} and in the limit of infinite dimensions~\cite{DMFT}.
But, when we have $J_{ij\downarrow} = J$ and $J_{ij\uparrow} = 0$, {\it i.~e.,~} if the electrons with spin up are held static while those with spin down are allowed to hop between nearest-neighbor sites, we obtain the Falicov-Kimball model, which can also be viewed as describing a mixture of heavy and light fermions on a lattice. This model has the advantage of displaying a Mott transition while also being more amenable to numerical methods than the Hubbard model. We will study this model at half-filling when there are as many electrons with spin up as spin down (or, equivalently, as many light, $\downarrow$, as heavy, $\uparrow$, fermions) and, in total, as many electrons as there are lattice sites.

\begin{figure}[t] %[htbp]
\begin{center}
{\includegraphics[width=14.0cm, height=5.50cm]{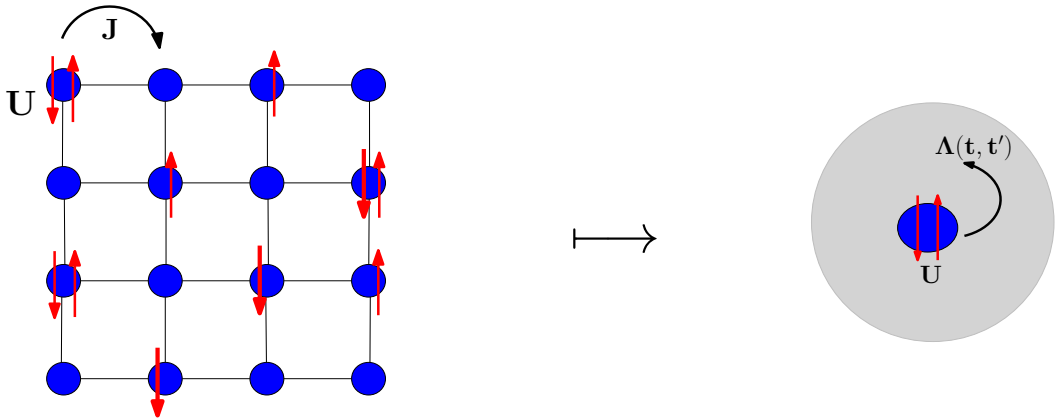}}
\caption{The dynamical mean-field theory maps the lattice problem (here, a square lattice with hoping integral $J$ between nearest-neighboring sites and Coulomb interaction $U$ for doubly occupied sites)  onto that of an impurity embedded into a self-consistently determined bath with a hybridization function $\Lambda(t,t')$. In equilibrium, the hybridization function only depends on the relative time $t-t'$.
}  
\label{fig:lattice2DMFT}
\end{center}
\end{figure}

\section{Nonequilibrium DMFT}
\label{sec:NonEqDMFT}
Dynamical mean-field theory was introduced to address strongly correlated systems in equilibrium and has been used successfully to describe key aspects of strongly correlated systems such as the metal-to-Mott insulator transition~\cite{DMFT, DMFT_2, DMFT_3, DMFT_FK}.
The method has successfully been adapted to studies of nonequilibrium systems~\cite{DMFT_noneq, FK_NonEq_DMFT08, DMFT_noneq_Aoki}. It maps the lattice model onto an impurity embedded in a self-consistently determined bath as illustrated in figure~\ref{fig:lattice2DMFT}. The method then relies on an accurate solution of the impurity problem and has been implemented with a multitude of impurity solvers with varied levels of success. These solvers include diagrammatic perturbative approaches~\cite{pertubationTheory_NonEq, pertubationTheory_NonEq_2}, Quantum Monte Carlo~\cite{CTQMC_NonEq} and exact diagonalization~\cite{KrylovPotthoff, Arrigoni_von_der_linden}. The Falicov-Kimball model considered in this case has the advantage of being able to be solved exactly, as shown below.
\vspace{10pt}

\subsection{Transient Nonequilibrium DMFT} 
Here we consider a system on a hypercubic lattice in the limit of infinite spatial dimensions ($d \to \infty $) with a constant electric field $\mathbf{E}$ applied at a specific time and oriented along a diagonal of the lattice. The dynamical mean-field theory employs a local self-energy such that
$$\Sigma_{k\sigma}(t,t') = \Sigma_{\sigma}(t,t'). $$
In the case of a system initially in equilibrium at an initial temperature $T=1/\beta$, the symbol $\langle \mathcal{A}(t)\mathcal{B}(t') \rangle$ in Eqs.~(\ref{eq:GLesser} - \ref{eq:GAdvanced}) is synonymous with $\mathrm{Tr}\, \mathrm{e}^{-\beta \mathcal{H}_{eq} }\mathcal{A}(t)\mathcal{B}(t') / \mathcal{Z}_{eq}$, where $\mathcal{H}_{eq}=\mathcal{H}(t_{\rm min})$ is the initial Hamiltonian and $ \mathcal{Z}_{eq} =  {\rm Tr}\,\mathrm{e}^{-\beta \mathcal{H}_{eq} }$ is the corresponding partition function. 

The contour-ordered Green's function obeys the Dyson equation given by
\begin{equation}
G_{k\sigma} (t, t') = G^0_{k\sigma}(t, t') 
+ \int_c d\bar{t} \;\int_c d\bar{t'} G^0_{k\sigma}(t,\bar{t}) \Sigma_{\sigma}(\bar{t} ,\bar{t'})G_{k\sigma} (\bar{t'}, t'),
\label{eq:FullG}
\end{equation}
where $G^0_{k\sigma}(t,t')$ is the noninteracting ($U{=}0$) Green's function [also defined by Eqs.~(\ref{eq:GLesser} - \ref{eq:GAdvanced}), but with a Hamiltonian that has $U=0$]. The integrals each range over the entire Kadanoff-Baym-Keldysh contour.
Prior to the electric field being turned on, the noninteracting Hamiltonian is
\begin{equation}
\mathcal{H}_0 = \sum_{k\sigma} ( \epsilon_k - \mu ) c_{ k\sigma }^{\dagger}c_{k \sigma }^{\phantom\dagger}.
\label{eq:HamiltonianNonInt}
\end{equation}
Here, $\epsilon_k$ is the band structure for the lattice electrons.
When the electrons move on a hypercubic lattice in infinite dimensions, the band energy is given by:
\begin{equation}
\epsilon_k = - \lim_{d \to \infty} \frac{t^*}{\sqrt {d}} \sum_{i=1}^{d} \mathrm{cos}( k_i a),
\end{equation}
where $a$ is the lattice spacing (and is set equal to 1); the nearest-neighbor hopping is rescaled via $J = t^*/2\sqrt{d}$ and $t^*$ is the unit of energy. In the paramagnetic phase, the spin index may be dropped.

The system is driven out of equilibrium by an applied electric field 
$E(r,t)$ which can be expressed in terms of a scalar potential  $\phi(r, t)$ and a vector potential $A(r,t)$ via
\begin{equation}
E(r,t) = - \nabla \Phi \left( r, t \right) - \frac{1}{c} \frac{ \partial A(r, t) }{\partial t}.
\end{equation}
We choose the Hamiltonian gauge ($\Phi = 0$), so that the electric field is completely determined by the vector potential. The electric field then enters into the 
Hamiltonian via the Peierls' substitution which modifies the hopping amplitude with a time-dependent phase~\cite{Peierls}:
\begin{equation}
J_{i j} \to J_{ij}\, \mathrm{exp} \left[ -\frac{i e}{ \hbar c} \int_{R_i}^{R_j} A(r, t) dr \right].
\end{equation}
We will be working with a spatially uniform, but time-varying electric field, which is given by a spatially uniform vector potential $A(t)$. In this case, the problem remains translationally invariant, but note that a spatially uniform but time-varying electric field does mildly violate Maxwell's equations. We ignore those magnetic-field effects since they are small.

The noninteracting Green's function obeys the equation:
\begin{equation}
 \left[i\partial_t + \mu - \epsilon_{k-\frac{eA(t)}{\hbar c}} \right] G^0_{k\sigma}(t,t') = \delta_c(t,t'), 
\end{equation}
where $\delta_c(t,t')$ is a generalization  of the Dirac delta function onto the contour.

When the field is constant and is turned on at time $t = 0$, the vector potential becomes $A( t) = - cEt\theta(t)$ and the Peierls' substitution results in the transformation $k \to k- \frac{e A(t)}{\hbar c} = k + \frac{e E t\theta(t)}{\hbar}$ in the Hamiltonian. 
The noninteracting Green’s function then depends on the band energy $\epsilon_k$ and on the band velocity $\bar{\epsilon_k} = - \lim_{d \to \infty} \frac{t^*}{\sqrt {d}} \sum_{i=1}^{d} \mathrm{sin}( k_i a)$ in the direction of the electric field. Setting $\hbar = 1 $ and $c=1$, this leads to the following expression~\cite{noninteracting_Turkowski}:
\begin{equation}
G^{0}_{k\sigma}(t, t')  = -i \left[ \mathrm{\theta}_c(t, t') - f(\epsilon_k -\mu) \right] 
\mathrm{exp}\left[ -i \int_{t'}^t d\bar{t}\left( \epsilon_{k + e E \bar{t}\mathrm{\theta}(\bar{t}) } - \mu \right) \right],
\end{equation}
which becomes
\begin{eqnarray}
G^{0}_{k\sigma}(t, t') &=& -i \left[ \theta_c(t, t') - f(\epsilon_k -\mu) \right]  \mathrm{e}^{i \mu (t-t') } \nonumber \\
&\times& \mathrm{exp}\left[  -i\int_{t'}^t d\bar{t} \left\{ \left. \bigg ( \mathrm{\theta}(-\bar{t})   + \mathrm{\theta}(\bar{t}) \mathrm{cos} \left( e E \bar{t} \right) \right. \bigg )\epsilon - \mathrm{\theta}(\bar{t}) \bar{\epsilon}\, \mathrm{sin} \left( e E \bar{t} \right) \right. \bigg \} \right. \Bigg ].
\label{eq: NonIntG}
\end{eqnarray}

The dressed lattice Green's function is constructed from the noninteracting Green's function and the self-energy following Eq.~(\ref{eq:FullG}). The local Green's function $G_\sigma(t,t')$ is found by summing the momentum-dependent dressed Green's function $G_{k\sigma}(t,t')$ over all momentum, or, equivalently integrating over the joint density of states, $\rho_2(\epsilon,\bar{\epsilon})$. 
 In the nonequilibrium DMFT formalism, the hybridization of the impurity to the dynamical mean field is given by
\begin{equation}
 \Lambda(t,t')  =  \left( i\partial_t + \mu \right) \delta_c(t,t') -  G^{-1}(t,t') + \Sigma(t,t'),
 \label{eq:hybridization}
\end{equation}
where we suppressed the spin indices.

To complete the self-consistency loop, we calculate the local Green's function from the dynamical mean field and use Dyson's equation (on the contour) to extract the self-energy. This is usually the bottleneck in the DMFT formalism. But in the case of the Falicov-Kimball model, this step can be determined analytically. We start from the spinless Falicov-Kimball model in the absence of a field
\begin{equation}
{\mathcal H}_{eq}  =  -  \frac{1}{2\sqrt{d}} \sum_{\langle ij\rangle} t^*_{ij} \left( c^{\dagger}_{i}c_{j}^{\phantom\dagger} + c^{\dagger}_{j}c_{i}^{\phantom\dagger}
 \right)  -  \mu \sum_{i}  c^{\dagger}_{i}c_{i}^{\phantom\dagger}  +  U\sum_{i}w_{i}c^{\dagger}_{i}c_{i}^{\phantom\dagger},
\label{eq:HamiltonianFK}
\end{equation}
where $w_{i}=0, 1$ is the occupation number operator of a localized fermions at site $i$. 
The corresponding impurity problem is solved by
\begin{eqnarray}
 G_{imp}(t,t') &=& (1-\langle w \rangle) \left[ \left( i\partial_t + \mu \right) \delta_c(t,t')  - \Lambda(t,t') \right]^{-1} \nonumber \\
               &+& \langle w \rangle \left[ \left( i\partial_t + \mu -U \right) \delta_c(t,t')  - \Lambda(t,t') \right]^{-1},
\label{eq:impurityG}               
\end{eqnarray}
where $\langle w \rangle = \sum_i \langle w_i \rangle /N$ is the initial (equilibrium) filling of the localized fermions. Note that the Green's function is represented by sum of the matrix inverses of two continuous matrix operators.\\

%\vspace{-1.0in}
\begin{figure}[htb]
\begin{center}
 \includegraphics*[width=7.50cm, height=7.50cm]{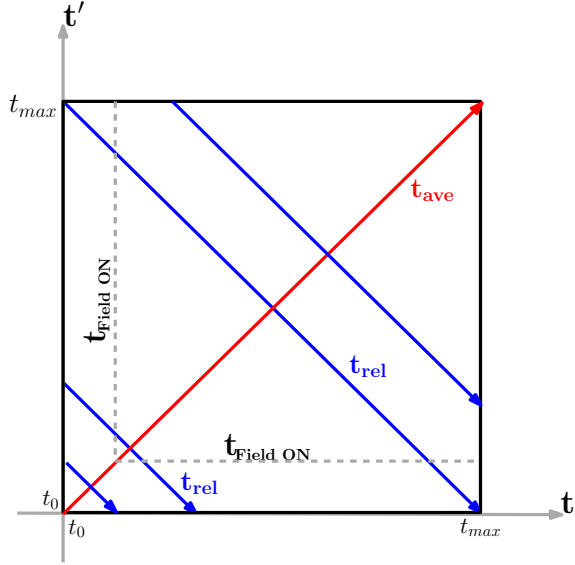}
\caption{Schematic representation of the transformation of time coordinates $(t,t')$ on the contour into the Wigner coordinates of average and relative time $\left [t_{\rm ave}=(t+t')/2,t_{\rm rel}=t-t'\right ]$. The dashed gray line represents the time at which the system is driven away from equilibrium by switching on the electric field (above and to the right of the dashed lines is where the field is turned on). The average time axis is illustrated by the red line across the diagonal and  each average time has an associated set of relative times identified by points running along the blue lines.
} 
\label{fig:KeldyshContour2WignerCoord}
\end{center}
\end{figure}
\vspace{0.50in}

\subsection{Nonequilibrium steady State DMFT}
\label{sec: SteadyState}

The main shortcoming of the above nonequilibrium DMFT solution is that even in cases such as that of the Falicov-Model, where we can write an exact expression for the impurity Green's function, the nonequilibrium DMFT implementation remains computationally expensive both in terms of memory and time (the compute-intensive step is determining the local Green's function from the self-energy). It is for this reason that most studies are limited to examining the short-time transient only. One may alternatively be interested in studying the system after it has undergone its relaxation from an initial state and settled into its steady state. In this case, a steady-state nonequilibrium DMFT formalism can be formulated directly to study the steady-state regime. In this situation, we can use a noninteracting initial state with its noninteracting contour-ordered Green's function given by:
\begin{equation}
G^{0}_k(t, t') = -i \theta(t-t') \mathrm{exp}\left[ -i \int_{t'}^t d\bar{t}\left( \epsilon_{k + \frac{e E \bar{t}}{\hbar}} - \mu \right) \right].
\end{equation}
Switching to Wigner coordinates, as illustrated in figure~\ref{fig:KeldyshContour2WignerCoord}, with the relative time $t_{\rm rel}=t-t'$ and average time $t_{\rm ave}=(t+t')/2$, we obtain:
\begin{eqnarray}
G^{0}_k(t, t') &=& -i \theta(t_{\rm rel}) \mathrm{exp}\left[ i \mu t_{\rm rel} -\frac{2i}{E t_{\rm rel}}\mathrm{sin} \left( \frac{E t_{\rm rel}}{2} \right)  \left\{ \epsilon_{k}\mathrm{cos} \left(E t_{\rm ave} \right) - \bar{\epsilon}_{k} \mathrm{sin} \left( E t_{\rm ave} \right) \right\} \right] 
\label{eq: NonIntG-Wigner}\\
&\equiv&g^0_k(t_{\rm rel}, t_{\rm ave})
\end{eqnarray}
in the long-time limit $(t\gg t_{\rm on}$ and $t'\gg t_{\rm on}$, with $t_{\rm on}$ the time the field is initially turned on).
We denote the Green's function in Wigner coordinates by $g^0_k(t_{\rm rel}, t_{\rm ave})$.
From Eq.~(\ref{eq: NonIntG-Wigner}), it is easy to show that 
\begin{equation}
 G^0_{k - E\tau} (t, t') = G^0_k(t - \tau, t' - \tau)
 \label{eq: GreenSym1}
\end{equation}
or, equivalently
\begin{equation}
\label{eq: GreenSym2}
 g^0_{k - E\tau} (t_{\rm rel}, t_{\rm ave}) = g^0_k(t_{\rm rel},t_{\rm ave} - \tau),
\end{equation}
when the system is driven by a DC field.

Using the steady state assumption (in the long-time limit) that $ \Sigma (t, t') = \Sigma(t - t')$, one obtains that the dressed Green's functions also satisfy the expressions in Eqs.~(\ref{eq: GreenSym1})  and (\ref{eq: GreenSym2}).
%$ G_{k - E\tau} (t, t') = G_k(t - \tau, t' - \tau) $ or $g_{k - E\tau} (t_{\rm rel}, t_s) = g_k(t_{\rm rel},t_s - 2 \tau)$.
For the choice of $\tau = (t + t')/2$, we obtain $G_{k - E\tau} (t, t') = G_k\left ((t - t')/2,(t' - t)/2\right )$. The local Green's function is $G(t, t') = \sum_k G_k ((t-t')/2, (t'-t)/2) $, which, in terms of the relative time, becomes
\begin{equation}
G(t, t') \equiv  g(t-t', t_{\rm ave}\to\infty)= \sum_k g_k ( t-t', t_{\rm ave}\to\infty).
\end{equation}
Note how $g_k$ depends on $k$ only through $\epsilon_k$ and $\bar{\epsilon}_k$ leading to
\begin{equation}
g(t-t',t_{\rm ave}\to\infty) =  \int_{ - \infty}^{+ \infty}\int_{ - \infty}^{+ \infty} d \epsilon \; d \bar{\epsilon} \;
\rho_2(\epsilon, \bar{\epsilon}) g_{\epsilon \bar{\epsilon}} (t-t', t_{\rm ave}\to\infty) 
\end{equation}
with the joint density of states~\cite{DMFT} given by
\begin{equation}
 \rho_2(\epsilon, \bar{\epsilon})  =  \frac{1}{\pi } \exp \left[ - \epsilon^2- 
 \bar{\epsilon}^2 \right].
\end{equation}
This means that in frequency space, the local Green's function is obtained from
\begin{equation}
g(\omega, t_{\rm ave}\to\infty) = \int_{ - \infty}^{+ \infty}\int_{ - \infty}^{+ \infty} d \epsilon \; d \bar{\epsilon} \;  
\rho_2(\epsilon, \bar{\epsilon}) g_{\epsilon \bar{\epsilon}} (\omega, t_{\rm ave}\to\infty) \equiv g(\omega). \label{eq:localG}
\end{equation}
One can similarly show that in the long-time limit, we have $G_k (t + \frac{2 \pi}{E}, t'+ \frac{2 \pi}{E}) = G_k(t, t')$ or equivalently, 
$ g_k(t_{\rm rel},t_{\rm ave} - \frac{2 \pi}{E}) = g_k(t_{\rm rel},t_{\rm ave}) $ {\it i.~e.,} $g_k$ is periodic with respect to $t_{\rm ave}$ with period $2\pi/E$; this is the Bloch-Zener periodicity and $\nu_n = nE$ (with
$n$ an integer) are the Bloch frequencies. There is a subtle issue going on here, because we are working in this long-time limit. The formulas stated here hold when both times are much larger than $t_{\rm on}$. Nevertheless, the Green's function has a periodic average time dependence in this long-time limit, and this is what we are describing here. As a result, Fourier transforming $g_k ( t_{\rm rel}, t_{\rm ave})$ in this long-time limit, gives
\begin{equation}
g_k ( \omega, \nu_l) = \frac{E}{2 \pi } \int d t_{\rm rel}\; \mathrm{e}^{ i \omega t_{\rm rel}}\int_0^{\frac{2 \pi }{E}} dt_{\rm ave} \; \mathrm{e}^{ i \nu_l t_{\rm ave}} g_k ( t_{\rm rel}, t_{\rm ave}) 
\end{equation}
where $\omega$ is a continuous variable while $\nu_l$ is quantized and the integral over $t_{\rm ave}$ can be restricted to be taken over just one Bloch period. In frequency space, the Dyson equation can then be written in the form
\begin{equation}
 g_k ( \omega, \nu_l) = g^0_k ( \omega, \nu_l)  +  \sum_m g^0_k ( \omega + \nu_m, \nu_l - \nu_m) \Sigma( \omega - \nu_l + 2\nu_m) g_k (\omega -\nu_l + 
\nu_m, \nu_m). \label{eq:Dyson_eq1}
\end{equation}
The periodicity in this long-time limit, maps the system to behave exactly as a many-body Floquet system acts. We now describe how to solve the steady-state problem within such a Floquet formalism, using discrete matrices within the Dyson equation for the Green's function defined within a frequency range determined by the Floquet period. We start by choosing a finite frequency range over which the Green's function has nonzero spectral weight.
Next, we choose an integer $L \gg 1$ such that $\Sigma(\omega) \ne 0$ only for $\omega \in \left[ -\nu_L, \nu_L \right]$ and let $\phi$ be a real variable allowed to vary continuously between $-E/2$ and $E/2$. Equation (\ref{eq:Dyson_eq1}) for $ (\omega, \nu_l)= (\phi+\nu_{-L+p}, \nu_p)$ can be rewritten as follows:
\begin{equation}
g_{ip}(\phi) = g^0_{ip}(\phi) + \sum_{j=0}^L A_{i j}(\phi) g_{jp}(\phi) 
= g^0_{ip}(\phi) + \sum_{j=0}^L g^0_{i j}(\phi)\Sigma_j(\phi) g_{jp}(\phi) \label{eq:Dyson_eq2} \end{equation}
and one can immediately see that the frequency $\phi$ is coupled only to frequencies shifted by integer multiples of the Bloch frequency.
Here we have defined
\begin{eqnarray}
\Sigma_i(\phi) &=& \Sigma(\phi+ \nu_{-L + 2i})\\
g^0_{i j}(\phi) & = & g^0(\phi+\nu_{-L + i+j}, \nu_{i -j} )\\
g_{i j}(\phi) & = & g(\phi+\nu_{-L + i+j}, \nu_{i -j} )\\
A_{i j}(\phi) & = & g^0_{i j} (\phi)\Sigma_j(\phi) 
\end{eqnarray}
with $i, j, p$ being integer indices. Equation~(\ref{eq:Dyson_eq2}) provides a convenient way to solve the Dyson equation. For this purpose, it is necessary 
to obtain the Fourier transform of the noninteracting Green's function in Eq.~(\ref{eq: NonIntG-Wigner}) to frequency space. 
The result is
\begin{equation}
g^0_{\epsilon \bar{\epsilon}}( \omega, \nu_n) = R(\omega, \nu_n) + D(\omega, \nu_n)
\end{equation}
with
\begin{eqnarray}
&~&R( \omega, \nu_n) = \frac{ \pi e^{-i n \alpha}}{E \; \mathrm{sin} \left(\pi u - \frac{\pi |n|}{2}\right)} J_{\frac{|n|}{2} - u}(r') J_{\frac{|n|}{2} + u}(r')  \\
&~&~~= \frac{2 e^{-i n \alpha}}{E} \sum_{m = -\infty}^{+\infty} \mathcal{P}\left (\frac{1}{2u -m}\right ) J_{\frac{1}{2}|m| - \frac{1}{2}|n|}(r' )   J_{\frac{1}{2}|m| + \frac{1}{2}|n|}(r' ) \left\{ \begin{array}{l l}
\mathrm{cos}^2 \left( \frac{\pi n}{2} \right) & \quad \textrm{if $m = 2k$} \\
\mathrm{sgn}(m) \sin^2 \left( \frac{\pi n}{2} \right)& \quad \textrm{if $m = 2k+1$} \\
\end{array} \right. \nonumber
\end{eqnarray}
and
\begin{eqnarray}
D(\omega, \nu_n) &=& -2\pi i \sum_{m = -\infty}^{+\infty} \delta(2u - m) e^{-i n \alpha} \nonumber \\
&\times& J_{\frac{1}{2}|m| - \frac{1}{2}|n|}(r ) J_{\frac{1}{2}|m| + \frac{1}{2}|n|}(r ) \left\{ \begin{array}{l l}
\mathrm{cos}^2 \left( \frac{\pi n}{2} \right) & \quad \textrm{if $m = 2k$} \\
\mathrm{sgn}(m) \mathrm{sin}^2 \left( \frac{\pi n}{2} \right)& \quad \textrm{if $m = 2k+1$} \\
\end{array} \right. 
\end{eqnarray}
where $J_a(x)$ is the Bessel function of the first kind and $\mathcal{P}$ denotes the principal value. The different variables are defined by
$ u = \frac{\omega + \mu}{E}$, $r = \sqrt{\epsilon^2 + \bar{\epsilon}^2}$, 
$r' = r/E$, $\mathrm{cos} \alpha = \frac{\epsilon}{\sqrt{\epsilon^2 + \bar{\epsilon}^2}}$ and
$\mathrm{sin} \alpha = \frac{\bar{\epsilon}}{\sqrt{\epsilon^2 + \bar{\epsilon}^2}}$.

Note that the integer indices $i, j$ and $p$ in Eq.~(\ref{eq:Dyson_eq2}), are chosen such that $i, j, p = 0, 1, 2, \cdots L$, for the quantized frequencies $\nu_{i}$, $\nu_j$, and $\nu_p$, which combine with the real variable $\phi \in [-E/2, E/2]$ to produce a dependence on a single continuous frequency that varies within the range of nonzero spectral weight $[-\nu_L, \nu_L]$. This single variable is relabelled as $\omega$. 

To complete the self-consistency loop for this nonequilibrium steady-state DMFT algorithm, the only missing ingredient is the impurity solver. 
In the case of the Falicov-Kimball model [which is initially described by the Hamiltonian in Eq.~(\ref{eq:HamiltonianFK})], the dressed Green's function is related to the noninteracting Green's function via
\begin{equation}
g(\omega) = (1-w_1)g^0(\omega) + \frac{w_1}{g^{0^{-1}}(\omega) - U}.
\end{equation}
Combining this with the Dyson equation yields
\begin{equation}
\label{eq:sigma_loc}
\Sigma(\omega) = \frac{ \left[ 1 + g(\omega) \Sigma(\omega)\right] U w_1}{1 + g(\omega)\left[ \Sigma(\omega) -U \left(1 - w_1 \right) \right]}
\end{equation}
The self-consistency loop is now complete and proceeds as follows: 
\begin{enumerate}
 \item For a given $\Sigma(\omega)$, solve the Dyson equation for $g_{\epsilon,\bar{\epsilon}}(\omega, t_{\rm ave}\to\infty)$;
 \item Next, solve Eq.~(\ref{eq:localG}) for the local Green's function;
 \item Then, solve Eq.~(\ref{eq:sigma_loc}) for $\Sigma(\omega)$;
 \item Repeat steps $1-3$ until convergence is achieved.
\end{enumerate}

This formalism is equivalent to using Floquet theory to solve the steady-state problem for a field driven system~\cite{ Floquet_Shirley, Floquet_Sambe, steadyStateJoura_2, steadyState_Aoki}. It allows the characterization of the steady state for a given electric field and interaction strengths. In particular, one can obtain the local density of states (DOS) from $\rho(\omega) = - \mathrm{Im}[g(\omega)]/\pi$.

\begin{figure}[htbp]
\begin{center}

\includegraphics*[width=16.0cm, height=4.0cm]{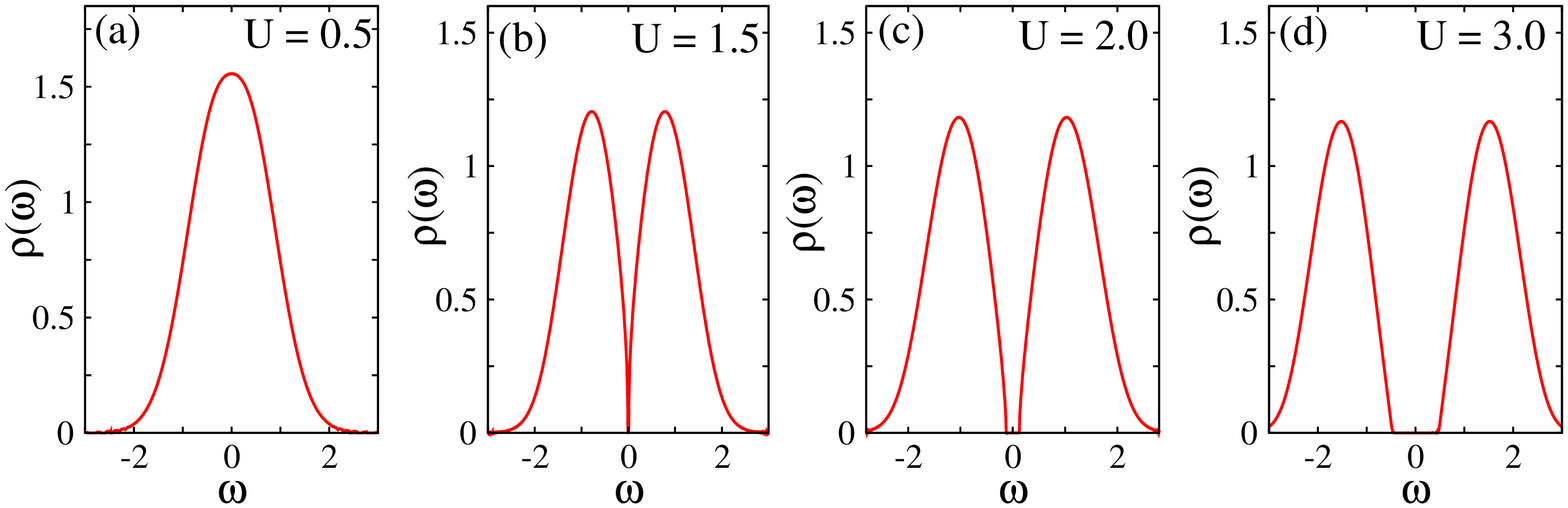}
\caption{Local density of states of the Falicov-Kimball model solution in equilibrium for $U = 0.5$ (a), $U = 1.50$ (b),  $U = 2.0$ (c), $U = 3.0$ (d). The system develops a Mott-insulating gap as the interaction is increased with the transition occurring when $U=\sqrt{2}$.
} 
\label{fig:Eq_DOS_4Us}
\end{center}
\end{figure}

\section{Relaxation Dynamics and Fluctuation Dissipation Theorem}
\label{sec:NonEqDynamicsResults}

\subsection{Steady-state density of states}

In order to be able to describe the effects of the electric field, it is instructive to start by looking at the equilibrium local density of states (DOS) in figure~(\ref{fig:Eq_DOS_4Us}) obtained by the DMFT algorithm in equilibrium~\cite{DMFT_FK}.
As the interaction strength is increased, we see a transition from a broadened Gaussian-like DOS in the weak-interaction regime to a gapped spectrum with two peaks in the strong-interaction regime. The spectral gap at $\omega=0$ is characteristic of the Mott-insulating regime, with the critical $U$ occurring at $U=\sqrt{2}$. Note that the DOS of the Falicov-Kimball model, in the normal state, does not depend on temperature.

\begin{figure}[htbp]
\begin{center}
\includegraphics*[width=16.0cm, height=4.0cm]{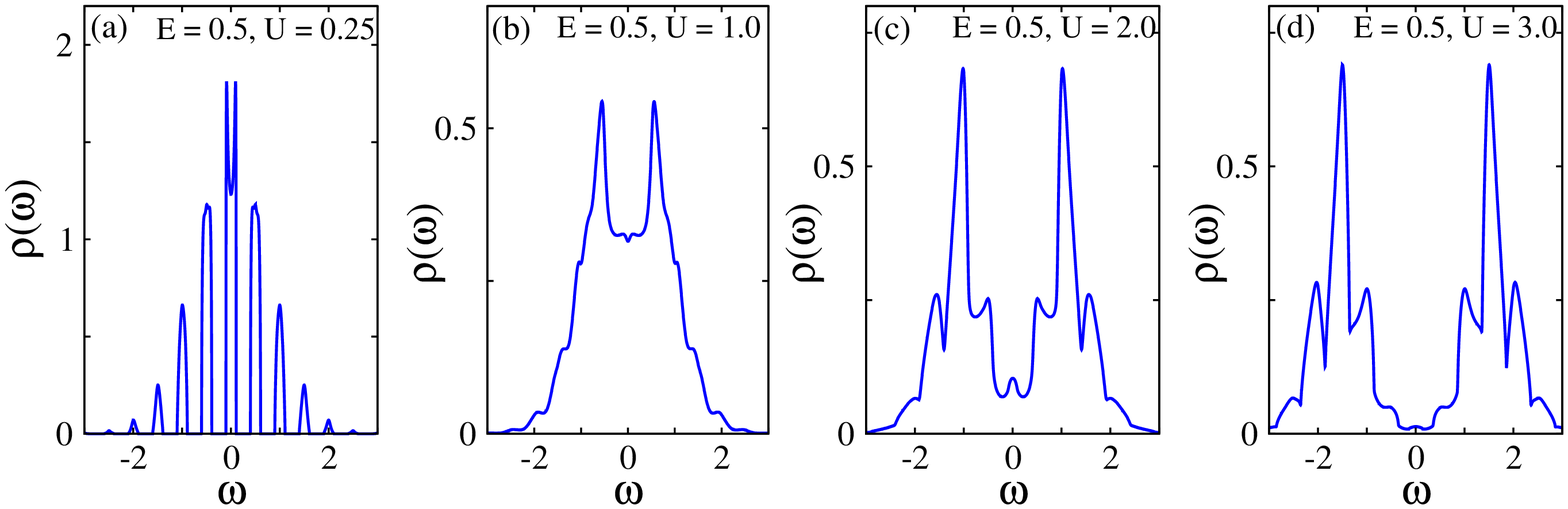}
\caption{Local density of states of the field-driven system in its steady state for $E = 0.5$ and $U = 0.25$ (a),  $U = 1.0$ (b), $U = 2.0$ (c), $U = 3.0$ (d).  
} 
\label{fig:NonEqSS_DOS_E0p5_4Us}
\end{center}
\end{figure}

\begin{figure}[htbp]
\begin{center}
\includegraphics*[width=16.0cm, height=4.0cm]{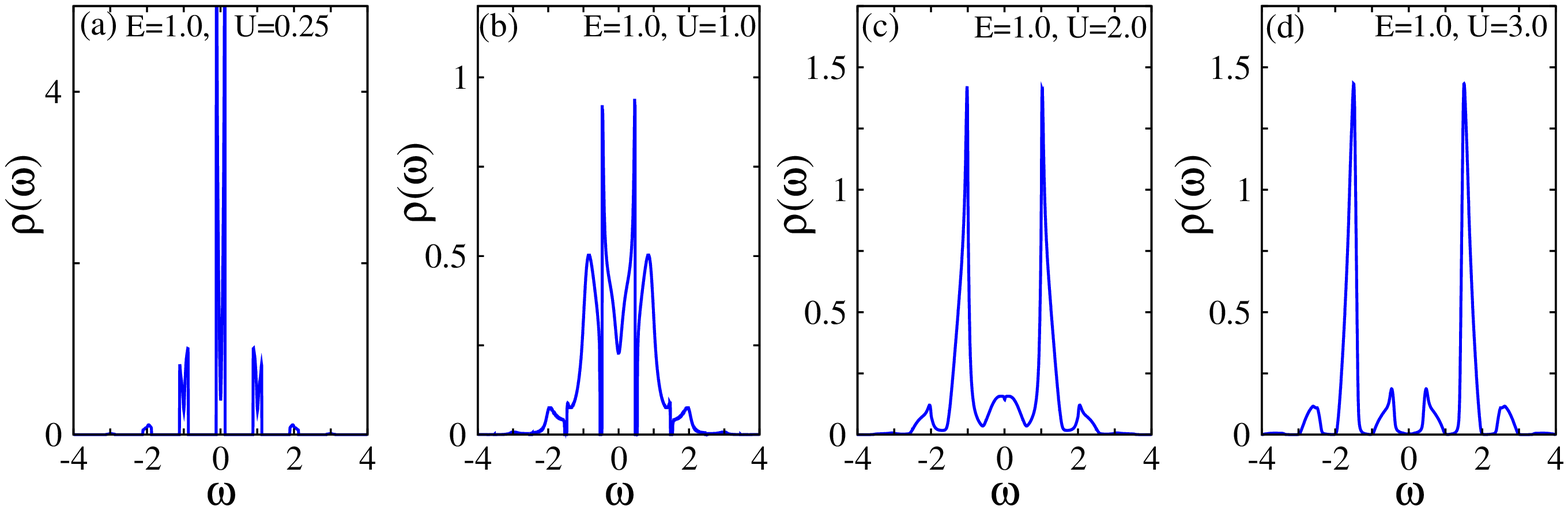}
\caption{Local density of states of the field-driven system in its steady state for $E = 1.0$ and $U = 0.25$ (a),  $U = 1.0$ (b), $U = 2.0$ (c), $U = 3.0$ (d).
} 
\label{fig:NonEqSS_DOS_E1p0_4Us}
\end{center}
\end{figure}

The nonequilibrium steady state DOS shows much richer behavior than in equilibrium. Figures \ref{fig:NonEqSS_DOS_E0p5_4Us}  and \ref{fig:NonEqSS_DOS_E1p0_4Us} show the steady-state DOS for electric fields $E = 0.5$ and $E = 1.0$, respectively, for the same interaction strengths as the equilibrium DOS in figure~\ref{fig:Eq_DOS_4Us}. In the weak-interaction regime, the DOS displays the Wannier-Stark ladder~\cite{Wannier} with peaks centered around $\omega = nE$ with $n = 0, \pm 1, \pm 2, ...$  and the amplitude is sharply suppressed away from the central peak. The peaks are broadened by a width governed by the interaction strength and the central peak has a dip separated by two side peaks at $\pm U/2$. As  the interaction is further increased, the initial peaks eventually merge, while preserving the central dip that arises from strong correlations. With stronger interactions this dip does not however become the full width of the gap due to the presence of ``in gap'' states created by the electric field.

In both the equilibrium and the steady-state regime, the system obeys the fluctuation-dissipation theorem and the lesser Green's function satisfy the relation
\begin{equation}
 g^<(\omega) = -2 i f_T(\omega) \mathrm{Im}\left[ g^R(\omega) \right].
 \label{eq:FDT_Glesser_Gretarded}
\end{equation}
Where $f_T(\omega)$ is the Fermi distribution function at temperature $T$. It is interesting to now ask how do the dynamics transiently take us from equilibrium to the steady state?

\begin{figure}[htbp]
\begin{center}
\includegraphics*[width=8.0cm, height=7.0cm]{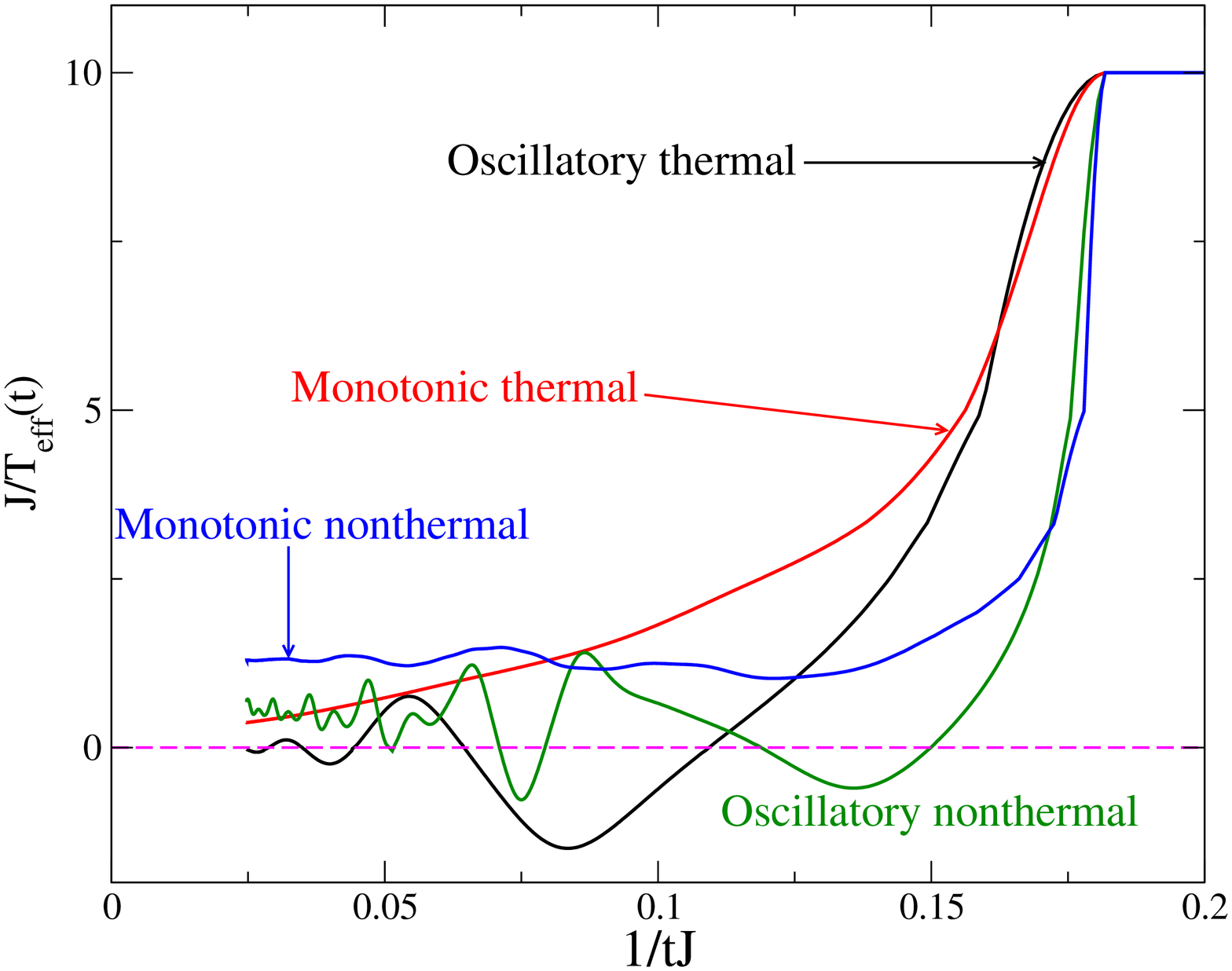}
\caption{Effective temperature $T_{\rm eff}$ as a function of time for the field-driven system. The dashed magenta line corresponds to infinite temperature and the data is plotted so that the vertical axis shows the inverse of the effective temperature while the horizontal axis displays the inverse of time. Along with the temperature, the dynamics of several physical quantities are used to identify different relaxation scenarios by examining different parameters (different field strength and interaction values). We observe monotonic thermal (red), oscillatory thermal (black), monotonic nonthermal (blue) and oscillatory nonthermal (green) relaxation. Nonthermal relaxation is characterized by not going to the infinite-temperature limit at long times. Reprinted figure with permission from H. F. Fotso, K. Mikelsons and J. K. Freericks, Scientific Reports {\bf 4}, 4699 (2014).
} 
\label{fig:effectiveT_VS_t}
\end{center}
\end{figure}

\subsection{Monotonic thermalization}

In earlier work, we examined the transient dynamics of correlated quantum systems when they are driven away from equilibrium by a DC electric field ~\cite{thermalization}. Key findings included the fact that the DOS is only constrained by causality and takes its steady-state value as soon as the field is turned on (seeing this in the frequency domain takes a little longer, because one needs the retarded Green's function behavior to hold for a long-enough relative time span that a Fourier transformation can be performed). Since the system is isolated and the electric field remains constant after it is switched on,  Joule heating should lead to an infinite-temperature state, if the system thermalizes. Without assuming the system to be thermal, an effective temperature at a time $t$ can be obtained by comparing the energy of the system to that of the corresponding equilibrium system (in other words, using a thermometer based of the energy as a scale, without determining if the system has actually thermalized). The evolution of the effective temperature as a function of time is presented in figure~\ref{fig:effectiveT_VS_t}. We found that for some parameter regimes, this infinite-heating scenario did take place. Sometimes the infinite-temperature limit was approached monotonically, sometimes it was approached in an oscillatory fashion, with the effective temperature alternating between positive and negative temperatures en route to $T=\infty$.

\begin{figure}[htbp]
\begin{center}
\includegraphics*[width=13.50cm, height=7.50cm]{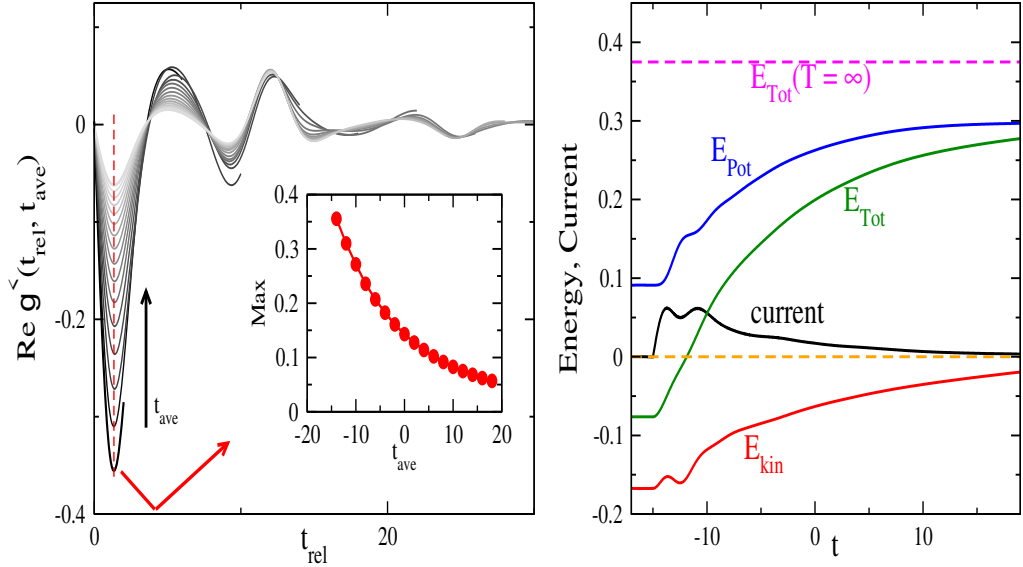}
\caption{
Physical indication of the monotonic evolution of the system towards an infinite temperature thermal state for $E = 0.5$ and $U = 1.5$ evolving from a system initially in equilibrium at temperature $T = 0.1$. The left panel shows the real part of the lesser Green’s function as a function of $t_{\rm rel}$ for successive average times (the field is turned on at $t_{\rm on}=-15$); a grayscale is used with lighter shades indicating later average times. The inset shows the maximum (or absolute value of the minimum) of $g^<(t_{\rm rel})$ as a function of average time. The panel on the right shows the total energy ($E_{\mathrm{Tot}}$, green), the potential energy ($E_{\mathrm{Pot}}$, blue), kinetic energy ($E_{\mathrm{Kin}}$ red) and the current (black) as functions of time. The dashed line indicates the total energy at infinite temperature for the same interaction strength. Reprinted figure with permission from H. F. Fotso, K. Mikelsons and J. K. Freericks, Scientific Reports {\bf 4}, 4699 (2014).
} 
\label{fig:MonotonicThermalization_E0p5_U1p5}
\end{center}
\end{figure}

We focus on the case of monotonic relaxation towards the infinite-temperature thermal state as it exhibits interesting manifestations of the fluctuation-dissipation theorem. Besides the effective temperature from an energy thermometer, several other quantities are used in characterizing the relaxation. These include the real part of the frequency-dependent lesser Green's function (figure~\ref{fig:MonotonicThermalization_E0p5_U1p5}, left panel), and the current, the kinetic, potential and total energy (figure~\ref{fig:MonotonicThermalization_E0p5_U1p5}, right panel). If the system is in a thermal state, then the lesser Green's function and the retarded Green's function satisfy the fluctuation-dissipation theorem. Namely, in frequency space, they obey Eq.~(\ref{eq:FDT_Glesser_Gretarded}).

\begin{figure}[htbp]
\begin{center}
\includegraphics*[width=10.0cm, height=8.0cm]{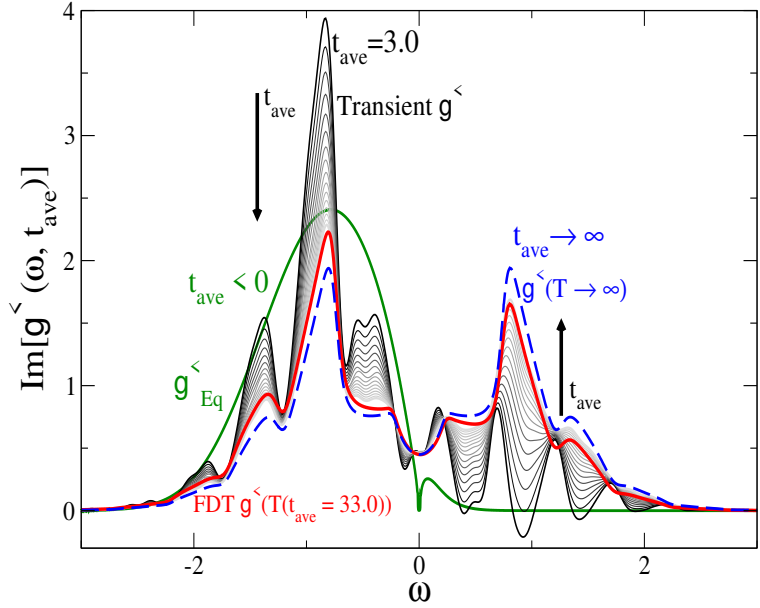}
\caption{Imaginary part of the lesser Green's function as a function of frequency for initial temperature $T = 0.1$, $U = 1.5$ and $E = 0.5$ corresponding to a monotonic, thermalized case. The green line indicates the equilibrium curve before the electric field is turned on. The evolution of the nonequilibrium lesser Green's function, after the early transient, is shown by the grayscale plots with lighter shades indicating later average times. As the lesser Green's function evolves towards its infinite-temperature steady-state value (blue dashed curve), it is shown to be well matched by the fluctuation-dissipation-theorem result, after an early transient. This is illustrated with the fluctuation-dissipation theorem curve at the latest average time (red curve) for a system at the corresponding effective temperature (as determined by the energy thermometer). While not perfectly matching the fluctuation-dissipation theorem result, it is quite close. Reprinted figure with permission from H. F. Fotso, K. Mikelsons and J. K. Freericks, Scientific Reports {\bf 4}, 4699 (2014).
} 
\label{fig:FDT_GlesserW_EqTransient}
\end{center}
\end{figure}

In this case, the lesser Green's function has a vanishing real part both in time  and in frequency space. Figure~\ref{fig:MonotonicThermalization_E0p5_U1p5}, left panel, shows that the real part of the lesser Green's function relaxes monotonically towards zero (its infinite-temperature value). Additionally, it is easy to evaluate the total energy at infinite temperature.
This is given by the dashed magenta line in figure~\ref{fig:MonotonicThermalization_E0p5_U1p5}, right panel. To evaluate the transient total energy of the system, one starts from the known equilibrium energy at the initial temperature $T = 0.1$. Next, we integrate the inner product $\langle \mathbf{j}(t) \rangle \cdot \mathbf{E}$ up to time $t$ to obtain the energy added to the system by Joule heating (here $\mathbf{j}(t)$ is the current at time $t$). Adding this Joule heating energy to the initial energy produces the total energy of the system at time $t$. In figure~\ref{fig:MonotonicThermalization_E0p5_U1p5}, one can see that in this case the total energy in the transient calculation approaches this limit monotonically, as the current also decays monotonically towards zero.

This decay of the current towards zero is to be expected at infinite temperature where particles are equally likely to move in opposite directions. The potential and kinetic energy also monotonically approach their infinite-temperature values. In this monotonic thermalization scenario, we find that even before reaching the infinite-temperature thermal state, the system appears to follow an evolution through successive quasi-thermal states approximately satisfying the fluctuation-dissipation theorem. This is illustrated in figure \ref{fig:FDT_GlesserW_EqTransient} for the lesser Green's function as a function of frequency and average time. After the field is turned on, $g^<$ switches from its equilibrium value (green line) to different transient values. Their evolution is pictured in grayscale with lighter shades indicating later average times. At infinite average time, $g^<$ is expected to adopt the infinite-temperature fluctuation-dissipation result (blue dashed line). It is seen in the graph that at the last available  simulation times, the transient already follows the fluctuation-dissipation theorem: the red line is the fluctuation-dissipation theorem at the effective temperature (using energy as the thermometer scale) corresponding to this average time and shows good agreement with the transient.

\begin{figure}[htbp]
\begin{center}
\includegraphics*[width=7.50cm, height=7.0cm]{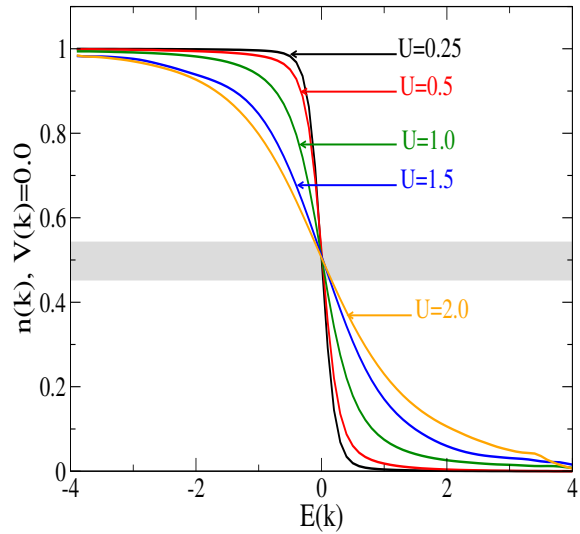}
\caption{
Plot of $n_{\mathbf k}$ as a function of $E({\mathbf k})$ for $V({\mathbf k})=0.0$ and for different values of $U$; the shaded area represents 
the range $[0.45, 0.55]$ over which $n_{\mathbf k}$ is plotted in figures \ref{fig:equilibriumFSurface} and \ref{fig:WignerDistSnapshots}. The deviation from the Fermi-Dirac distribution in equilibrium for $T=0.1$ comes from the many-body effects of the interactions between the two types of atoms. Reprinted figure with permission from H. F. Fotso, J. C. Vicente and J. K. Freericks,  Phys. Rev. A {\bf 90}, 053630 (2014). Copyright 2020 by the American Physical Society.
} 
\label{fig:FermiDistribution}
\end{center}
\end{figure}

Another window into the relaxation of the field driven system is through the distribution function~\cite{WignerPatternsPRA}. For the noninteracting system in equilibrium, this is simply the Fermi-Dirac distribution function. This equilibrium distribution function is modified by many-body effects for finite $U$ values although the lineshape is essentially preserved as shown in figure \ref{fig:FermiDistribution}. We envision this situation as corresponding to fermionic atoms of two different masses trapped in a mixture on an optical lattice~\cite{AtesZiegler}.

\begin{figure}[htbp]
\begin{center}
\includegraphics*[width=8.0cm, height=7.50cm]{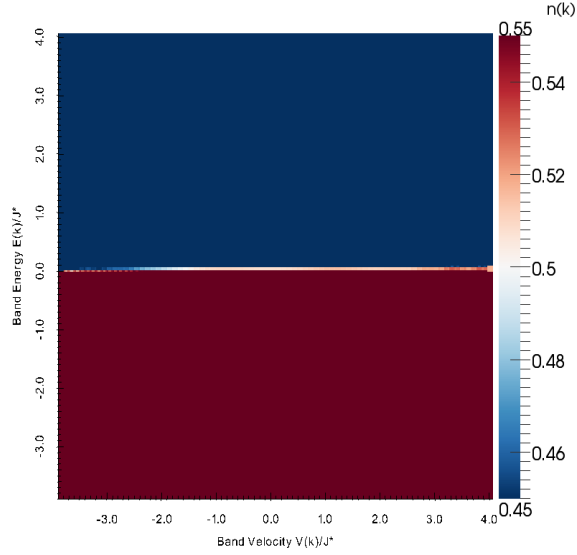}
\caption{
False-color image of the initial equilibrium Wigner distribution function at temperature $T=0.1$. The graph shows $n_{\mathbf k}$ as a function of $E({\mathbf k}) \equiv \epsilon_k$ and $V({\mathbf k}) \equiv \bar{\epsilon_k}$ for $U=0.25$. When the distribution function is larger than 0.55, it is plotted with the color at 0.55 and similarly when it is smaller than 0.45.  Reprinted figure with permission from H. F. Fotso, J. C. Vicente and J. K. Freericks,  Phys. Rev. A {\bf 90}, 053630 (2014). Copyright 2020 by the American Physical Society.
} 
\label{fig:equilibriumFSurface}
\end{center}
\end{figure}

For the interacting system away from equilibrium, the distribution function is given by the gauge-invariant Wigner distribution $n_{\mathbf k}(t) = -i G^<_{{\mathbf k}+{\mathbf A}(t)}(t, t)$~\cite{jauho}. Here, $n_{\mathbf k}(t)$ represents the occupation of states in momentum space. The dependence on $k$ is recast into a dependence on the band energy and the band velocity. Prior to the electric field being switched on, the system is in equilibrium at a temperature $T=0.1$ and the Wigner distribution in momentum space is shown in figure \ref{fig:equilibriumFSurface} for $U=0.25$ and is similar for other interaction strengths.

\begin{figure}[htbp]
\begin{center}
\includegraphics*[width=16.0cm, height=4.0cm]{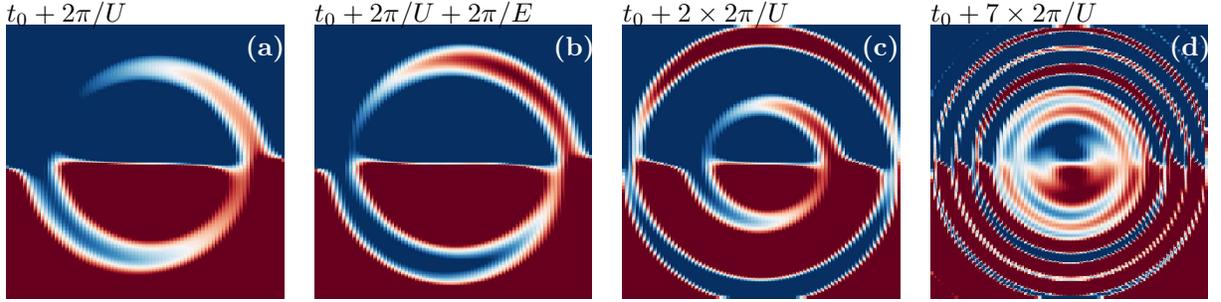}
\caption{False color snapshots of the evolution of the gauge-invariant Wigner distribution in momentum space at different times for $E=2.0$, and $U=0.25$~\cite{WignerPatternsPRA}. Each panel shows $n_{\mathbf k}(t)$ at an instant in time after the field is switched on (at $t_0$). New rings are formed on a timescale of $2\pi/U$ ((a), (c), (d)) and spiral around the origin on a $2\pi/E$ timescale (b).  When the distribution function is larger than 0.55, it is plotted with the color at 0.55 and similarly when it is smaller than 0.45. This is done to zoom in on the interesting patterns, which are a few percent effect. Reprinted figure with permission from H. F. Fotso, J. C. Vicente and J. K. Freericks,  Phys. Rev. A {\bf 90}, 053630 (2014). Copyright 2020 by the American Physical Society.
} 
\label{fig:WignerDistSnapshots}
\end{center}
\end{figure}

Once the field is turned on, it is expected that the onset of the electric current will lead to Joule heating that will increase the temperature until the system has reached an infinite-temperature state where all points in momentum space, or equivalently in [band energy:$E({\mathbf k}) \equiv \epsilon_k$; band velocity:$V({\mathbf k}) \equiv \bar{\epsilon_k}$]-space,  are equally likely to be occupied. In this state, the current vanishes and the Wigner distribution is now homogeneous for all $k$. This is translated into the false color plot being entirely white for the plotted region. It was previously reported that the evolution between the initial configuration of figure~\ref{fig:equilibriumFSurface} to this uniform configuration goes through the development of different patterns that depend on the field and interaction strength. For instance, for a strong electric field and weak interaction ($E=2.0, \; U = 0.25$), we observe in the false-color plots of the Wigner distribution as a function of time, the formation of ring-shaped disturbances on a timescale of $2\pi/U$ [figure \ref{fig:WignerDistSnapshots}-(a), (c), (d)]. These rings spiral around the center of the graph on a timescale of $2\pi/E$ [figure \ref{fig:WignerDistSnapshots}-(b)] and we see the formation of additional rings after every $2\pi/U$ time step [figure \ref{fig:WignerDistSnapshots}-(c),(d)]. $2\pi/E$ is the period of Bloch oscillations while $2\pi/U$ is the timescale for the collapse and revival of Bloch oscillations~\cite{poles, MGreinerBlochNature, BuchleitnerKolovsky}. As the interaction is increased ($E=2.0, \; U = 1.0$), these two timescales merge and we no longer observe well-separated individual rings but rather the growth of a single spiral that gradually scrambles the occupation of the states. We also observe at larger interaction strengths ($E=2.0, \; U = 3.0$) that long-lived features with both a stable region of high occupation and a region of low occupation  rotate around the origin at the Bloch period.\\

\section{Discussion}

In general, the fluctuation-dissipation theorem is not satisfied once a system is driven away from equilibrium. This review describes nonequilibrium DMFT studies for both the transient and the steady-state of the Falicov-Kimball model, describing a Fermi-Fermi mixture of heavy-light particles, when it is driven away from equilibrium by a constant electric field. We showed the complex range of relaxation scenarios exhibited by this nonequilibrium system. In particular, the density of states is fundamentally altered by the electric fields with the formation of Wannier-Stark ladders and a dielectric breakdown that arises with the presence of mid-gap states  that are absent in equilibrium. This density of states however switches to its steady-state value rapidly after the field is turned on. For an isolated system, the relaxation to a steady state satisfying the fluctuation-dissipation theorem can then occur in several identified scenarios. We have particularly examined the monotonic thermalization scenario where the system monotonically goes to an infinite-temperature thermal state (satisfying the fluctuation-dissipation theorem) and evolves through consecutive quasi-thermal states (approximately satisfying the fluctuation-dissipation theorem) en-route.  Ongoing work will take advantage of this property to develop an extrapolation scheme that bridges the gap between the transient and the steady state at minimal computational cost. We have also described how the key timescales that appear in the current in the form of Bloch oscillations and their collapse and revival, or beats, are manifested in the Wigner distribution function and its evolution towards infinite temperature where all states are equally occupied  [$n_{\mathbf k}=0.5$]. These results illustrate the rich physics of field-driven correlated quantum systems and the role the fluctuation-dissipation theorem plays in understanding this behavior.\\

\section*{Acknowledgments} 

This work was supported by the Department of Energy, Office of Basic Energy Sciences, Division of Materials Sciences and Engineering under Contract No. 
DE-FG02-08ER46542 (Georgetown). J.K.F. was also supported by the McDevitt bequest at Georgetown.


\begin{thebibliography}{99}
%--------------------------------
%--------------------------------

%Cold atoms
\bibitem{MGreinerBlochNature} M. Greiner, O. Mandel, T. W. H\"ansch and I. Bloch, Nature {\bf 419}, 51 (2002). %Collapse and revival of matter wave field

%Cold atoms
\bibitem{BlochDalibardZwerger} I. Bloch, J. Dalibard, and W. Zwerger, Rev. Mod. Phys. {\bf 80}, 885 (2008).

%Nanodevices, beyond linear response
\bibitem{MeirWingreen_PRL1992} Y. Meir and N.S. Wingreen, Phys. Rev. Lett. \textbf{68}, 2512 (1992).

\bibitem{CaroliNozieres_JPhysC1971} C. Caroli, R. Combescot, P. Nozieres and D. Saint-James, J. Phys. C: Sol. State Phys. \textbf{4}, 916 (1971).

%Pump probe
\bibitem{Perfetti_et_al} L. Perfetti, P. A. Loukakos, M. Lisowski, U. Bovensiepen, H. Berger, S. Biermann, P. S. Cornaglia, A. Georges, and M. Wolf, Phys. Rev. Lett. {\bf 97}, 067402 (2006).

%Thermalization general
\bibitem{Deutsch1991}  J. M. Deutsch, Phys. Rev. A {\bf 43}, 2046 (1991). %Thermalization

\bibitem{RigolNature2008} M. Rigol, V. Dunjko, and M. Olshanii, Nature {\bf 452}, 854 (2008). %Thermalization

\bibitem{Srednicki1994} M. Srednicki, Phys. Rev. E {\bf 50}, 888 (1994).  %Thermalization

\bibitem{thermalization} H. F. Fotso, K. Mikelsons and J. K. Freericks, Scientific Reports {\bf 4}, 4699 (2014).  %Thermalization


\bibitem{NonEqPhaseTransition1} M. Dean, Y. Cao, X. Liu, S. Wall, D. Zhu, R. Mankowsky, V. Thampy, X. M. Chen, J. G. Vale, D. Casa et al., Nat. Mater. {\bf 15}, 601 (2016).

\bibitem{NonEqPhaseTransition2} T. Ogasawara, M. Ashida, N. Motoyama, H. Eisaki, S. Uchida, Y. Tokura, H. Ghosh, A. Shukla, S. Mazumdar, and M. Kuwata-Gonokami
Phys. Rev. Lett. {\bf 85}, 2204 (2000).

\bibitem{DMFT_noneq}  J. K. Freericks, V. M. Turkowski, and V. Zlati\'c,  Phys. Rev. Lett. {\bf 97}, 266408 (2006).

\bibitem{DMFT} W. Metzner and D. Vollhardt, Phys. Rev. Lett. {\bf 62}, 324 (1989).

\bibitem{DMFT_2} Y. Kuramoto, Springer Series in Solid State Science Vol. 62, edited by T. Kasuya and T. Sao (Springer, 1985), p. 152.

\bibitem{DMFT_3} E. M\"{u}ller-Hartmann, Z. Phys. B {\bf 74}, 507 (1989).

\bibitem{DMFT_FK} J. K. Freericks and V. Zlati\'c, Rev. Mod. Phys. {\bf 75}, 1333 (2003). %equilibrium DMFT FK

\bibitem{FK_NonEq_DMFT08} J. K. Freericks, Phys. Rev. B {\bf 77}, 075109 (2008).

\bibitem{DMFT_noneq_Aoki} H. Aoki, N. Tsuji, M. Eckstein, M. Kollar, T. Oka and P. Werner, Rev. Mod. Phys. {\bf 86}, 779 (2014).

\bibitem{steadyStateJoura_1} J. K. Freericks and A. V. Joura, ``Nonequilibrium density of states and distribution functions for strongly correlated materials 
across the Mott transition'', in Electron transport in nanosystems, edited by Janez Bonca and Sergei Kruchinin (Springer, Berlin, 2008) pp. 219--236.

\bibitem{steadyStateJoura_2} A. V.Joura, J. K. Freericks, and Th. Pruschke, Phys. Rev. Lett. {\bf 101}, 196401 (2008).

\bibitem{steadyState_Aoki} N. Tsuji, T. Oka, and H. Aoki, Phys. Rev. B {\bf 78}, 235124 (2008). %Floquet? Yes.

\bibitem{steadyState_Camille_FDT} C. Aron, G. Kotliar, C. Weber, Phys. Rev. Lett. {\bf 108}, 086401 (2012). %FDT NESS DMFT 

\bibitem{FalicovKimball} L. M. Falicov and J. C. Kimball, Phys. Rev. Lett. {\bf 22}, 997 (1969).

\bibitem{AtesZiegler}C. Ates and K. Ziegler, Phys. Rev. A {\bf 71}, 063610 (2005).

\bibitem{Keldysh64_65} L. V. Keldysh, Zh. Eksp. Teor. Fiz. {\bf 47}, 1945 (1964) [Sov. Phys. JETP {\bf 20}, 1018 (1964)].

\bibitem{BaymKadanoff62} L. P. Kadanoff and G. Baym, \textit{ Quantum Statistical Mechanics} ( Benjamin, New York, 1962).

\bibitem{LiebWu_1968} E.H. Lieb and F.Y. Wu, Phys. Rev. Lett. \textbf{20}, 1445 (1968).

\bibitem{pertubationTheory_NonEq} M. Eckstein and P. Werner, Phys. Rev. B {\bf 82}, 115115 (2010).

\bibitem{pertubationTheory_NonEq_2} P. Schmidt and H. Monien, preprint cond-mat {\bf 0202046} (2002).

\bibitem{CTQMC_NonEq} M. Eckstein, M. Kollar, and P. Werner,, Phys. Rev. Lett. {\bf 103}, 056403 (2009).

%\bibitem{tDMRG} R. Gezzi, T. Pruschke, and V. Meden, Phys. Rev. B {\bf 75}, 045324 (2007).

\bibitem{KrylovPotthoff} M. Balzer, N. Gdaniec and M. Potthoff, J Phys.: Cond. Matt. {\bf 24}, 3 (2012).

\bibitem{Arrigoni_von_der_linden} E. Arrigoni, M. Knap, and W. von der Linden, Phys. Rev. Lett. {\bf 110}, 086403  (2013).

\bibitem{noninteracting_Turkowski} V. M. Turkowski and J. K. Freericks, Phys. Rev. B \textbf{71}, 085104 (2005).


%\bibitem{tDMRG_2} S. White, and A. Feiguin,  Phys. Rev. Lett. 93, 076401 (2004).

%\bibitem{tDMRG_3} A. Daley, C. Kollath, U. Schollwock, and G. Vidal, J. Stat. Mech.: Theor. Exp , P04005 (2004).

\bibitem{Peierls} R. Peierls, Z. Phys. {\bf 80}, 763 (1933).

\bibitem{Floquet_Shirley} J. H. Shirley, Phys. Rev. {\bf 138}, B979 (1965).

\bibitem{Floquet_Sambe} H. Sambe, Phys. Rev. A {\bf 7}, 2203 (1973).

\bibitem{Wannier} G. H. Wannier, Phys. Rev. {\bf 100}, 1227 (1955); {\bf 101}, 1835 (1956);  {\bf 117}, 432 (1960); Rev. Mod. Phys. {\bf 34}, 645 (1962).

\bibitem{WignerPatternsPRA} H. F. Fotso, J. C. Vicente and J. K. Freericks,  Phys. Rev. A {\bf 90}, 053630 (2014).

\bibitem{jauho} R. Bertoncini and A. P. Jauho, Phys. Rev. B {\bf 44}, 3655 (1991).

\bibitem{Kolovsky} A. R. Kolovsky, Phys. Rev. Lett. {\bf 90}, 213002 (2003).%New Bloch Period for Interacting Cold Atoms in 1D Optical Lattices

\bibitem{MeinertNagerl} F. Meinert, M. J. Mark, E. Kirilov, K. Lauber, P. Weinmann, M. Gr\"obner and H.-C. N\"agerl, Phys. Rev. Lett. {\bf 112}, 193003 (2014). %Collapse and revival experiment 

\bibitem{poles} M. Mierzejewski, J.  Bon\v ca, and P. Prelov\v sek, Phys. Rev. Lett. {\bf 107}, 126601 (2011). %Field Driven Mott insulator


\bibitem{BuchleitnerKolovsky} A. Buchleitner and A. R. Kolovsky, Phys. Rev. Lett. {\bf 91}, 253002 (2003). %Interaction-Induced Decoherence of Atomic Bloch Oscillations


\end{thebibliography}
\end{document}